\documentclass[final,3p,times,twocolumn]{elsarticle}

\def\bea{\begin{eqnarray}}
\def\eea{\end{eqnarray}}

\newcommand{\bra}[1]{\langle #1|}
\newcommand{\ket}[1]{|#1\rangle}

\def\be{\begin{equation}}
\def\ee{\end{equation}}
\def\ba{\begin{eqnarray}}
\def\ea{\end{eqnarray}}

\usepackage{slashed}

\usepackage{graphics}
\usepackage{graphicx}
\usepackage{epsf}
\usepackage{amsmath}
\usepackage{amssymb}

\def\sfrac#1#2{{\textstyle \frac{#1}{#2}}}

\journal{Physics Letters B}

\begin{document}

\begin{frontmatter}



\title
{Using baryon octet magnetic moments and masses to fix the pion cloud contribution }

\author{Franz Gross$^{1,2}$, G. Ramalho$^{1,3}$, and K. Tsushima$^4$ \vspace{-0.05in} }

\address{
$^1$Thomas Jefferson National Accelerator Facility, Newport News,
VA 23606, USA \vspace{-0.05in}}
\address{$^2$College of William and Mary, Williamsburg, VA 23185, USA \vspace{-0.05in}}
\address{$^3$Centro de F{\'\i}sica Te\'orica de Part{\'\i}culas,
IST, Lisboa, Portugal\vspace{-0.05in} }
\address{$^4$EBAC in Theory Center,
Thomas Jefferson National Accelerator Facility,
Newport News,
VA 23606, USA  \vspace{-0.35in}}



\begin{abstract}
Using SU(3) symmetry to constrain the $\pi BB'$ couplings, assuming SU(3) breaking comes only from one-loop pion cloud contributions, and using the the covariant spectator theory to describe the photon coupling to the quark core, we show how the experimental masses and magnetic moments of the baryon octet can be used to set a model independent constraint on the strength of the pion cloud contributions to the octet, and hence the nucleon, form factors at $Q^2=0$.
\end{abstract}


\end{frontmatter}

The introduction of SU(3) as a symmetry of
strong interaction provided a systematic organization
of the low-lying baryon multiplets and also a simple
estimate for the baryon octet (containing the $N$, $\Lambda$, $\Sigma$, and $\Xi$ ground state baryons)  magnetic moments
in terms of two independent constants associated with the SU(3) symmetry \cite{GellMann62,Okubo62,Coleman61,Swart63}.
The extension to the SU(6) quark model \cite{Beg64}
predicted an empirically known relation $\mu_n/\mu_p=-2/3$,
but the description of the octet baryon magnetic moments
were still qualitative.
The explicit SU(3) flavor symmetry breaking
describing the octet baryon magnetic moments
as the sum of the three independent quark
magnetic moments (additive quark model)
that scaled with the inverse of the
respective quark mass, improved the
description of the octet magnetic moments
to a precision of 0.22$\mu_N$ \cite{PDG}
($\mu_N= \sfrac{e}{2m}$ is the nuclear magneton, with $m$ the nucleon mass).
We refer to these models (below) as naive quark models (NQM).

The NQM description
can be further improved with the addition of
meson cloud corrections motivated by chiral symmetry.
Good examples are the
Cloudy Bag Model (CBM)
\cite{Theberge83,Tsushima,Leinweber99},
and models that combine constituent quarks
with chiral phenomenology \cite{Barik86,Ha98,Cloet03,Faessler06}.
Alternatively, meson cloud contributions to the octet magnetic moments  can be calculated from models that use hadronic degrees of freedom constrained by chiral perturbation theory ($\chi$PT), chiral effective field theory ($\chi$EFT), or Heavy Baryon $\chi$PT \cite{Jenkins93,Puglia00,Geng09}.
From this perspective the loop
corrections involving heavy mesons or
intermediate heavy baryons (from the decuplet) are suppressed.
In practice however the convergence of
the chiral expansion is slow \cite{Geng09}
and/or involves a fit of several low energy constants
leading to contributions that are very
scheme dependent \cite{Puglia00,Geng09}.
The octet magnetic moments
have also been studied in lattice QCD \cite{Lattice}.

In this letter we show that rather simple assumptions about the  structure of the baryon octet can be used to fix the size of the pion cloud contributions to the octet electromagnetic form factors at $Q^2=0$.  As a consequence, the size of previously undetermined pion cloud contributions to the nucleon form factors are fixed and it is possible to assess the many models of the pion cloud that are currently in use.

In order to obtain these results we assume that the baryons are composed of three dressed valence quarks with an intrinsic structure that gives rise to quark form factors and quark anomalous moments.  We also assume that the coupling of the pion to the baryon octet is fixed by SU(3), and that the pion cloud contributions to the baryon masses and magnetic moments are well described by the lowest order one-loop contributions.   This latter assumption is supported not only by leading order chiral perturbation theory, but also by general arguments from dispersion theory that suggest that intermediate states with the lowest threshold are the most important.  This means that contributions from heavy mesons or two-pion loops should be smaller than the one-pion loop contributions we are considering.  Subject to these general assumptions, the results we present are {\it model independent\/}; they do not depend on the details of the dynamics.

\begin{figure}
\centerline{
\mbox{
\includegraphics[width=2.2in]{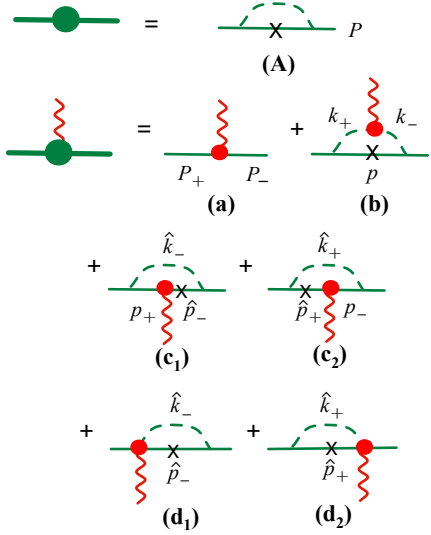}
}
}
\caption{\footnotesize\baselineskip=10pt (Color on line) Feynman diagrams (with time flowing from right to left) for the contributions to the octet self-energy (A) and the octet form factors (a) - (d) coming from a pion with $\gamma^\mu\gamma^5$ coupling.  These are written in for the CST with the internal baryons labeled by the $\times$ on-shell.}
\label{fig1}
\end{figure}

First, consider the baryon mass corrections that come from the pion self-energy loop shown in Fig.~\ref{fig1}(A).   Since $(\slashed{P})^2=P^2$ (where $P$ is the baryon four momentum), the self energy $\Sigma$ can be expressed as a function of $\slashed{P}$, and near a baryon pole at $P^2=M_B^2$, the baryon propagator has the form \cite{book}
\bea
\Delta(\not\!P)=\frac{1}{M_0-\not\! P+\Sigma(\not\! P)}=Z_B\bigg\{\frac{M_B+\not\!P}{M_B^2-P^2} + {\cal R}\bigg\},
\eea
where $M_0$ is the undressed mass of the baryon, ${\cal R}$ is finite at the pole at $P^2=M_B^2$, and
\bea
M_B&=&M_0+\Sigma_0 \nonumber\\
Z_B^{-1}&=&1-\Sigma'_0=1-\frac{d\Sigma(\not\!P)}{d\!\not\!P}\Big|_{\;\not P=M_B}\, , \label{eq:3}
\eea
with $\Sigma_0=\Sigma(M_B)$.  The self energy diagram at the pole can be written $\Sigma_0=G_{0B}B_0$, where $G_{0B}$ is a factor depending on the coupling of the pion to the baryon $B$, and $B_0$ is the value of the Feynman integral (with the couplings removed) at the mass $M_B$; we assume this integral to be very weakly dependent on the baryon mass and approximately the same for all baryons in the octet.  Note that the couplings $G_{0B}$,  through their dependence on the quantum numbers of the baryons (including strangeness),  account indirectly for the differences in the masses of the strange and non-strange constituent quarks \cite{Cloet:2002eg,Young:2009zb}.    Using this picture the size of $\Sigma$ depends primarily on the SU(3) dependence of the  $\pi B B'$ coupling constants, summarized in Table \ref{tab:1} \cite{GellMann62,Carruthers}.  Note the definition of the quantities $\beta_\Sigma=4(1-\alpha)^2$, $\beta_\Lambda=\sfrac43\alpha^2$, and $\beta_\Xi=(1-2\alpha)^2$, to be used later.
Incorporating the overall factor of $g^2$ into the definition of $B_0$ gives the mass splittings summarized in Table \ref{tab:2}.  Adjusting the three parameters $M_0$, $B_0$, and $\alpha$ to give a best fit to the average masses $M_N=939.0, M_\Lambda=1115.7, M_\Sigma=1192.4,$ and $M_\Xi=1318.1$ (in MeV) gives the parameters $M_0=1393.2,  B_0=-148.9,$ (in MeV) and $\alpha=0.6943$.  Only the value of $\alpha$ will be needed below; note that it is close to the 0.6 expected from SU(6).

\begin{table}
\begin{minipage}{3in}
\begin{tabular}{l c c}
\hline
\hline
$\pi B B'\qquad$ & ${\cal O}_{\pi B B'}$ & $\qquad g_{\pi B B'}\qquad$ \\
\hline
$\pi N N$ & $g_{_{\pi NN}} (\xi^*_\pi\cdot\tau)$ & $ g$\\[0.05in]
$\pi \Xi \Xi$ &  $g_{_{\pi \Xi\Xi}} (\xi^*_\pi\cdot\tau)$& $g\,(1-2\alpha)$\\[0.05in]
$\pi\Lambda\Sigma$ &  $g_{_{\pi \Lambda\Sigma}} (\xi^*_\pi\cdot\xi_\Sigma)$  & $\sfrac2{\sqrt{3}}\,g\,\alpha$\\[0.05in]
$\pi\Sigma\Sigma$ &  $g_{_{\pi \Sigma\Sigma}} (\xi^*_\pi\cdot{\bf J})$ & $2 g\,(1-\alpha)$\\[0.05in]
\hline
\hline
\end{tabular}
\caption{Pion-baryon couplings in $SU(3)$.  Here $\xi_\pi$ and $\xi_\Sigma$ are the isospin-one polarization vectors of the $\pi$ and $\Sigma$, $\tau^i$ are the isospin-1/2 matrices, and ${\bf J}^i$ are the isospin-one matrices.  For the diagonal operators, the isospin wave function of the initial and final baryon have been suppressed. }
\label{tab:1}
\end{minipage}
\end{table}

\begin{table}
\begin{minipage}{3in}
\begin{tabular}{l l }
\hline
\hline
$B\quad$ & $G_{0 B}$ \\
\hline
$N$ & $\sum_\lambda(\xi_{\pi\lambda}\cdot\tau)(\xi^*_{\pi\lambda}\cdot\tau)= 3$  \\
$\Xi$ &  $3(1-2\alpha)^2\sum_\lambda(\xi_{\pi\lambda}\cdot\tau)(\xi^*_{\pi\lambda}\cdot\tau)=3\beta_\Xi$   \\
$\Lambda$ &  $\frac43\alpha^2\sum_{\lambda\mu}(\xi_{\pi\lambda}\cdot\xi_{\Sigma\mu})(\xi^*_{\pi\lambda}\cdot\xi^*_{\Sigma\mu})=4\alpha^2=3\beta_\Lambda$   \\
$\Sigma$ &  $4(1-\alpha)^2\sum_\lambda(\xi_{\pi\lambda}\cdot{\bf J})(\xi^*_{\pi\lambda}\cdot{\bf J})$
\\&
$+\sfrac43\alpha^2\sum_\lambda(\xi_{\pi\lambda}\cdot\xi_{\Sigma\mu})(\xi^*_{\pi\lambda}\cdot\xi^*_{\Sigma\mu})
=2\beta_\Sigma+\beta_\Lambda$    \\[0.05in]
\hline
\hline
\end{tabular}
\caption{ Contributions of the one pion-loop to the baryon self energies.  Here $\alpha=D/(F+D)$ is the SU(3) mixing parameter.}
\label{tab:2}
\end{minipage}
\end{table}

Next, turn to the electromagnetic form factors of the octet. 
In the language of the covariant spectator theory (CST), the interaction of a photon with the octet, at the one-pion loop level, is the sum of the six Feynman diagrams shown in Fig.~\ref{fig1}(a)-(d).  Briefly, these are  (a) interaction with the quark core (denoted  $J_{0B}^\mu$), (b) interaction with the pion (denoted $J_\pi^\mu$), and finally the sum of the four diagrams (c$_1$), (c$_2$), (d$_1$) and (d$_2$), all describing the interactions of the quark core dressed by the pion bubble and denoted collectively by $J_{\gamma B}^\mu$.  Explicitly, the total current is
\be
J_B^\mu=
Z_B \left[
J_{0B}^\mu + J_\pi^\mu + J_{\gamma B}^\mu
\right], \label{eq:3a}
\ee
where $Z_B$ is the renormalization constant of Eq.~(\ref{eq:3}) which preserves the baryon charge (discussed below).

As the octet members are spin 1/2 particles
their current $J_B^\mu$ can be written
\be
J_B^\mu = e\Bigg\{F_{1B}(Q^2) \gamma^\mu + F_{2B}(Q^2)
\frac{i \sigma^{\mu \nu} q_\mu}{2M_B}\Bigg\},
\label{eqJgen}
\ee
where $F_{1B}(0)=e_B$ is the baryon charge (in units of the proton charge $e$) 
and $F_{2B}(0)=\kappa_B$ the baryon anomalous moment.  The spin-flip matrix element of this current is proportional to the magnetic moment.  For example, defining the baryon spinor in the usual way
\be
u({\bf P},s)=\sqrt{\frac{E_P+M_B}{2M_B}}
\left(
\begin{array}{c}
 1   \\
 \frac{\sigma\cdot{\bf P}}{E_P+M_B}
\end{array}
\right)\chi_s\, ,
\ee
where $E_B=\sqrt{M_B^2+{\bf P}^2}$ and $\chi_s$ is the two-component spinor with spin projection $s$ in the $\hat z$ direction, the spin-flip matrix element of the current in  the Breit frame with
${\bf q}=q_z {\bf \hat z}$ in the $z$ direction is
\be
\lim_{q_z\to0}\;\frac1{q_z}\bar u(\sfrac12{\bf q},+)\,J_B^x\, u(-\sfrac12{\bf q},-)=(e_B+\kappa_B)\frac{e}{2M_B}\, , \label{jx}
\ee
showing that $\mu_B=e_B+\kappa_B$ is the magnetic moment,
in natural baryon units of $e/(2M_B)$.
Experimental magnetic moments are reported
of nuclear magnetons; conversion to these units gives  $\mu_B=(e_B+\kappa_B)\sfrac{m}{M_B}$.


To describe the interaction with the quark core,
corresponding to the current $J_{0B}^\mu$, we use the
CST quark model 
introduced in Refs.~\cite{Nucleon,Omega}.
In this model a baryon is a system of
three constituent quarks, one of which is off-shell.  In the leading approximation the photon couples to the off-shell quark, leaving the two on-shell quarks (which are treated as a single on-shell ``di-quark'' with a fixed effective mass) to be spectators.
The expression for the core current becomes
\begin{align}
J_{0B}^\mu &= 3e \sum_{\Lambda}\int_k \overline{\Psi}_B (P_+,k)j_{q}^\mu \Psi_B(P_-,k)
\nonumber\\
&\to
e\left\{e_B
\gamma^\mu +
\kappa_{0B}
\frac{ i \sigma^{\mu \nu} q_\nu}{2M_B}
\right\}\;\;({\rm as}\, q\to0), \label{eqJ0}
\end{align}
where $P_+$ ($P_-$) are the final (initial) baryon momenta, respectively,  $k$ is the diquark momentum,  $\Psi_B$ is the baryon wave function with quark 3 off-shell (by convention), $e_B$ is the charge of the baryon,
and $\kappa_{0B}$ is the bare (i.e. undressed by the pion cloud)
anomalous moment of the baryon.  The total wave function is symmetric; contributions from terms with quarks 1 and 2 off-shell equal that from quark 3 off-shell, and are accounted for by the factor of 3. 
 The covariant integral is over the three momentum $k$ of the spectator diquark (its fourth component fixed by the on-shell condition, see Ref.~\cite{Nucleon})
and the sum is over the four states $\Lambda=\{s,\lambda\}$ of the diquark, where $s$ is the scalar (spin 0) state of the diquark, and $\lambda=\{0,\pm\}$ are the three polarization states of the vector diquark.
The current operator for the off-shell quark (3, by convention) is
\begin{align}
j^\mu_q=&j_{1q}(Q^2)\gamma^\mu+j_{2q}(Q^2)\frac{i\sigma^{\mu\nu}q_\nu}{2m}
\nonumber\\
\to&\,
e_{q}\gamma^\mu+\left(\kappa_q\frac{M_B}{m}\right)
\frac{i\sigma^{\mu\nu}q_\nu}{2M_B}\;\;({\rm as}\, q\to0),
\label{quarkc}
\end{align}
where the second line displays the consequence of defining the quark anomalous moments in nuclear magnetons, as we have done in previous work.
The functions $j_{iq}$ ($i=1,2$) define
the Dirac and Pauli {\it constituent quark\/} form factors
as operators acting in the quark flavor state $q=u,d,s$.
The explicit form for $j_{iq}$ was defined in Ref.~\cite{Omega}.

\begin{table*}[t]
\begin{center}
\begin{tabular}{l l l r r l l}
\hline
\hline
$B$   & $\qquad\ket{M_S}$  &   $\qquad\ket{M_A}$ &  $j_1^S$ & $j_1^A$ & $j_2^S$  & $j_2^A$  \\[0.05in]
\hline
p     &   $\sfrac{1}{\sqrt{3}} \left[\frac1{\sqrt{2}}
       (ud + du) u - \sqrt{2} uu d \right]$ &
       $\sfrac{1}{\sqrt{2}} (ud -du) u$  & 0 & $\sfrac23$
     &  $\sfrac{2}{9}(\kappa_u-\kappa_d)$ &
$\sfrac{2}{3}\kappa_u$
\\[0.05in]
n     &  $-\sfrac{1}{\sqrt{3}} \left[\frac1{\sqrt{2}}
        (ud + du) d - \sqrt{2} ddu \right]$ &
        $\sfrac{1}{\sqrt{2}} (ud -du) d$
        & $\sfrac13$ & $-\sfrac13$  &
$\sfrac{1}{9}(4 \kappa_u -\kappa_d)$ &
$-\sfrac{1}{3}\kappa_d$
\\[0.1in]
\hline
$\Sigma^+$  & $\sfrac{1}{\sqrt{3}} \left[\frac1{\sqrt{2}}(us + su) u - \sqrt{2} uu s \right]$ &
            $\sfrac{1}{\sqrt{2}} (us -su) u$ &  $0$ & $\sfrac23$  &
$\sfrac{2}{9}(\kappa_u-\kappa_s)$ & $\sfrac{2}{3}\kappa_u$
\\[0.1in]
$\Sigma^0$ &
$\sfrac{1}{\sqrt{12}}
\Big[(sd+ds)u$
&
$\sfrac{1}{2}\left[ (ds-sd)u+(us-su)d\right]$
& $-\sfrac16$  & $\sfrac16$  &
$\sfrac{1}{18}(2 \kappa_u -\kappa_d- 4 \kappa_s)$ &
$\sfrac{1}{3} (\kappa_u-\sfrac12 \kappa_d)$
\\
& $\quad+ (su+us)d -2(ud+du)s
\Big]$ && & &
\\[0.05in]
$\Sigma^-$ & $\sfrac{1}{\sqrt{3}}\left[\frac1{\sqrt{2}} (sd + ds) d - \sqrt{2} dd s \right]$ &
             $\sfrac{1}{\sqrt{2}} (ds -sd) d$
             & $-\sfrac13$  & $-\sfrac13$  &
$-\sfrac{1}{9}(\kappa_d + 2\kappa_s)$ & $-\sfrac{1}{3}\kappa_d$
            \\[0.1in]
\hline\\[-0.15in]
$\Lambda^0$ &
$\sfrac{1}{2}\left[ (ds+sd)u-(us+su)d\right]$
&
$\sfrac{1}{\sqrt{12}}\Big[(sd-ds)u$ & $\sfrac16$ & $-\sfrac16$ &
$\sfrac{1}{6}(2 \kappa_u -\kappa_d)$ &
$\sfrac{1}{18}(2 \kappa_u -\kappa_d $ 
\\
&&$-(su-us)d +2(ud-du)s
\Big]$&& & & $\qquad -4 \kappa_s)$
\\[0.05in]
\hline
$\Xi^0$ & $-\sfrac{1}{\sqrt{3}} \left[\frac1{\sqrt{2}}(us + su) s - \sqrt{2} ss u\right]$ &
         $\sfrac{1}{\sqrt{2}} (us -su) s$  &  $\sfrac13$ & $-\sfrac13$ &
$\sfrac{1}{9}(4 \kappa_u -\kappa_s)$ & $-\sfrac{1}{3}\kappa_s$ \\
$\Xi^-$ & $-\sfrac{1}{\sqrt{3}} \left[\frac1{\sqrt{2}}(ds + sd) s - \sqrt{2} ss d\right]$ &
         $\sfrac{1}{\sqrt{2}} (ds -sd) s$
         &  $-\sfrac13$  & $-\sfrac13$   &
$-\sfrac{1}{9}(2\kappa_d +\kappa_s)$ &
$-\sfrac{1}{3}\kappa_s$  \\
\hline
\hline
\end{tabular}
\end{center}
\caption{Flavor wave functions for quarks 1 and 2 in symmetric and antisymmetric configurations  and the corresponding matrix elements $j^S$ and $j^A$.}
\label{tab:4}
\end{table*}

The quark wave functions of the baryon octet SU(3) generalizations of the nucleon S-state wave functions used in Ref.~\cite{Nucleon}:
\be
\Psi_B(P,k) =
\frac{1}{\sqrt{2}} \left\{
\phi_S^0 \ket{M_A} +
\phi_S^1 \ket{M_S}
\right\} \psi(P,k)
\label{eq:5}
\ee
where $\ket{M_A}$ and $\ket{M_S}$ are flavor  wave functions
antisymmetric and symmetric in quarks 1 and 2,
and $\phi_S^X$ ($X=0,1$) are wave functions with quarks 1 and 2 in a spin
0 state (antisymmetric) and a spin-1 (symmetric) state,
respectively.
The function $\psi(P,k)$ is a scalar wave function
of $(P-k)^2$ properly normalized to one.
The states $\ket{M_A}$ and $\ket{M_S}$ are presented
in table \ref{tab:4}.
As for the spin states in their nonrelativistic
form with their relativistic generalization, are \cite{Nucleon}:
\ba
& &
\phi_S^0 =
\frac{1}{\sqrt{2}}
\left( \uparrow \downarrow - \downarrow  \uparrow  \right) \chi_s \to u(P,s)
\nonumber\\
& &
\phi_S^1
= -\sfrac{1}{\sqrt{3}}\sigma\cdot\varepsilon^{*}\chi_s\to -\sfrac{1}{\sqrt{3}}\gamma_5\not\!\varepsilon^*u(P,s)
\ea
where $\varepsilon$ is a covariant spin-1 polarization
vector describing the spin of the (12) diquark pair,
in the fixed axis basis \cite{FixedAxis}, and the Dirac spinor
$u(P,s)$ carries the spin of the third spectator
quark (see Refs.~\cite{Nucleon,FixedAxis} for details).

Since the wave functions (\ref{eq:5}) are direct products of spin and
flavor wave functions, we can separate
the spin part from the flavor part using
the flavor matrix elements
of the quark current (\ref{quarkc}), $j^A$ and $j^S$,
defined by
\begin{align}
j_i^A&=
\bra{M_A} j_{iq} \ket{M_A} \quad \;\;
j_i^S=
\bra{M_S} j_{iq} \ket{M_S}\, .
\end{align}
The explicit form for the coefficients $j_i^A$ and $j_i^S$
are presented in table \ref{tab:4}.
The baryon charge $e_B$ and bare anomalous moment $\kappa_{0B}$ can
then be calculated using
the results of Ref.~\cite{Nucleon} with
convenient replacement of the coefficients $j_i$:
\begin{align}
e_B&=\sfrac32  \left(j_1^A + j_1^S \right)
\nonumber \\
\kappa_{0B}&=
\left(\sfrac{3}{2} j_2^A
-\sfrac{1}{2}  j_2^S\right)\frac{M_B}{m}
-2  j_1^S . \label{eq:7}
\end{align}
The factor of $M_B/m$ multiplying the quark anomalous moments comes from the fact that they have been defined in nuclear magnetons, requiring the conversion shown in Eq.~(\ref{quarkc}).

Next we look at the contributions from the pion loop diagrams
Figs.~\ref{fig1}(b)-(d).
These diagrams can be written in the following form
\ba
J_\pi^\mu &=&
e\left(
B_1 \gamma^\mu + B_2
\frac{ i \sigma^{\mu \nu} q_\nu}{2M_B}
\right) G_{\pi B} \label{eq:15}  \\
J_{\gamma B}^\mu &=&
e\left(
C_1 \gamma^\mu + C_2
\frac{ i \sigma^{\mu \nu} q_\nu}{2M_B}
\right) \,G_{eB} +
\nonumber \\
& &
e\left(
D_1 \gamma^\mu + D_2
\frac{ i \sigma^{\mu \nu} q_\nu}{2M_B}
\right) \, G_{\kappa B} ,
\label{eq:10}
\ea
where the coefficients $B_i$, $C_i$, and $D_i$ are assumed to be independent of the baryon mass, and the $G_{x B}$ are coefficients which depend on each family of baryons in the octet are calculated from the SU(3) couplings given in Table \ref{tab:1}.   The $B_i$ are the sum of the contributions of diagrams (b) and (d)$_i$ [which have the same SU(3) structure as (b)], the $C_i$ are the contributions of the charge term $e_B\gamma^\mu$ in the bare current (\ref{eqJ0}) contributing to diagrams (c)$_i$, and the $D_i$ are the contributions of the undressed anomalous moment terms proportional to $\kappa_{0B}$.

The use of the factor $1/(2M_B)$ to normalize the $\sigma^{\mu\nu}q_\nu$ terms in Eqs.~(\ref{eq:15}) and (\ref{eq:10}) can be justified by looking at the detailed Feynman integral for Fig.~\ref{fig1}(b).  Omitting the overall factor $G_{\pi B}$, this integral becomes
\begin{align}
\left<\,J_\pi^x\right>
&=e \,\bar u(\sfrac12{\bf q},+)\int_k
\frac{(P_++P_--2k)^x(M_B-\slashed{k})}{{\cal D}(k_+,k_-)}u(-\sfrac12{\bf q},-)
\nonumber\\
&=e\bigg\{\int_k \frac{2k_x^2}{{\cal D}(k_+,k_-)}\bigg\}\;\bar u(\sfrac12{\bf q},+)\gamma^x u(\sfrac12{\bf q},-)
\nonumber\\
&=(B_1+B_2)\frac{eq_z}{2M_B}\, ,
\end{align}
where ${\cal D}$ is a denominator that depends on the details of the pion propagators and the structure of the current (but depends only on $k_x^2$, leading to the simplification in the second line) and the integral over $k$ is the covariant CST volume integral used previously \cite{Nucleon}.  Our normalization has lead to a result consistent with Eq.~(\ref{eq:15}), and examination of the integral in curly brackets shows that it is only weakly dependent of the baryon mass, leading to the conclusion that the $B_i$ are also.
We assume that similar arguments work for the coefficients $C_i$ and $D_i$.

Returning to the calculation, we calculate the factors $G_{\pi B}$ by summing over all possible isospin states of the intermediate pions using Table \ref{tab:1} and the isospin structure for the coupling of the photon to the pion
\bea
j_\pi=-i(\xi^*_{\pi\lambda'}\times\xi_{\pi\lambda})_3,
\eea
where $\lambda$ ($\lambda'$) are the isospin polarizations of the incoming (outgoing) pion.   Absorbing the factor of $g^2$ into the $B$'s, the results are reported in Table \ref{tab:5}.

To compute the coefficients $G_{zB}$ (where $z$ is either $e$ or $\kappa$) it is convenient to introduce a general operator decomposition for the bare hadronic currents.   For the $N$ and $\Xi$ isospin doublets, we will use the standard isoscalar-isovector notation
\bea
z_B=\sfrac12(z^s_B+z^v_B\tau_3),
\eea
where $e^s_N=e^v_N=1, e^s_\Xi=-e^v_\Xi=-1$.
The $\Sigma$ are isovectors, which we will decompose into three states according to
\bea
z_\Sigma=z^0_\Sigma {\bf 1}+\sfrac12\Big(z^1_\Sigma {\bf J}_3+z^{2}_\Sigma {\bf J}^2_3\Big),
\eea
where ${\bf J}_3$ is the third component of the isospin one operator.  With this notation,
\bea
z^0_{\Sigma}&=&z_{_{\Sigma^0}}
\nonumber\\
z^1_{\Sigma}&=&z_{_{\Sigma^+}}-z_{_{\Sigma^-}}
\nonumber\\
z^{2}_{\Sigma}&=&z_{_{\Sigma^+}}+z_{_{\Sigma^-}}-2z_{_{\Sigma^0}}\, ,
\eea
where $e^0_\Sigma=e^2_\Sigma=0$ and $e^1_\Sigma=2$. The coefficients are computed in Table \ref{tab:6}.  Note that $e_\Lambda=0$.

\begin{table}
\begin{minipage}{3in}
\begin{tabular}{l l }
\hline
\hline
$B$ & $G_{\pi B}$ \\
\hline
$N$ & $-i\sum_{\lambda'\lambda}(\tau\cdot\xi_{\pi\lambda'})(\tau\cdot\xi^*_{\pi\lambda})(\xi^*_{\pi\lambda'}\times\xi_{\pi\lambda})_3=2\tau_3$
\\[0.05in]
$\Lambda$ &  $-\sfrac43i \,\alpha^2\sum_{\lambda'\lambda\,\mu}  (\xi_{\pi\lambda'} \cdot\xi^*_{\Sigma\mu})(\xi^*_{\pi\lambda} \cdot\xi_{\Sigma\mu})(\xi^*_{\pi\lambda'}\times\xi_{\pi\lambda})_3=0$
\\[0.05in]
$\Sigma$ &  $-4i(1-\alpha)^2\sum_{\lambda\lambda'}(\xi_{\pi\lambda'}\cdot{\bf J})(\xi^*_{\pi\lambda}\cdot{\bf J})(\xi^*_{\pi\lambda'}\times\xi_{\pi\lambda})_3$\\
&$-\sfrac43i\alpha^2\sum_{\lambda\lambda'}(\xi^*_{\Sigma\mu'}\cdot\xi_{\pi\lambda'})(\xi_{\Sigma\mu}\cdot\xi^*_{\pi\lambda})(\xi^*_{\pi\lambda'}\times\xi_{\pi\lambda})_3$    \\
&$ =\Big(4(1-\alpha)^2+\sfrac43\alpha^2\Big){\bf J}_3\equiv (\beta_\Sigma+\beta_\Lambda) \,{\bf J}_3$
\\[0.05in]
$\Xi$ &  $-(1-2\alpha)^2i\sum_{\lambda'\lambda}(\tau\cdot\xi_{\lambda'})(\tau\cdot\xi^*_{\lambda})(\xi^*_{\pi\lambda'}\times\xi_{\pi\lambda})_3$  \\
& $=2(1-2\alpha)^2\tau_3= 2\beta_\Xi\,\tau_3$\\
\hline
\hline
\end{tabular}
\caption{ Values of $G_{\pi B}$ computed from Table \ref{tab:1} and diagram \ref{fig1}(b).}
\label{tab:5}
\end{minipage}
\end{table}

\begin{table}
\begin{minipage}{3in}
\begin{tabular}{l l }
\hline
\hline
$B$ & $G_{z B}$ \\
\hline
$N$ & $\tau^i\frac12(z^s_N+z^v_N\tau_3)\tau^i=\sfrac12(3z_N^s-z_N^v\tau_3)$
\\[0.05in]
$\Lambda$ &  $\sfrac43 \,\alpha^2{\rm Tr}\bigg\{z^0_\Sigma {\bf 1}+\sfrac12\Big(z^1_\Sigma {\bf J}_3+z^{2}_\Sigma {\bf J}^2_3\Big)\bigg\}=\beta_\Lambda(3z^0_\Sigma+z^{2}_\Sigma)$
\\[0.05in]
$\Sigma$ &  $4 \,(1-\alpha)^2\,{\bf J}^{i}\bigg\{z^0_\Sigma{\bf 1}+\sfrac12\Big(z^1_\Sigma {\bf J}_3+z^{2}_\Sigma {\bf J}^2_3\Big)\bigg\} {\bf J}^{i}+ \sfrac43 \,\alpha^2\,z_{_{\Lambda}} {\bf 1}  $\\
&$=\Big(\beta_\Sigma (2z^0_{\Sigma}+z^2_{\Sigma})+\beta_\Lambda z_{\Lambda}\Big){\bf 1}+\sfrac12\,\beta_\Sigma\,(z^1_\Sigma{\bf J}_3 -z^2_\Sigma{\bf J}_3^2)$
\\[0.05in]
$\Xi$ &  $(1-2\alpha)^2\tau^i\frac12(z^s_\Xi+z^v_\Xi\tau_3)\tau^i=\beta_\Xi \sfrac12(3z_\Xi^s-z_\Xi^v\tau_3)$\\[0.05in]
\hline
\hline
\end{tabular}
\caption{ Values of $G_{z B}$ computed from Table \ref{tab:1} and diagram \ref{fig1}(b).  In each expression, the sum over pion polarizations has been carried out.}
\label{tab:6}
\end{minipage}
\end{table}

\begin{table*}
\begin{center}
\begin{tabular}{l c r r  r r r r c r}
\hline
\hline\\[-0.1in]
$B$   & $\quad \mu_{0B}= (e_B+ \kappa_{0B})\frac{m}{M_B}$  &
$\mu_{0B}^{{\rm no}\,\pi}$
& $\quad\mu_{B}^{\rm NQM}$ 
& $\mu_{B}^{\rm CBM}$  & $ Z_B\,\mu_{0B}$ & $Z_B\,\delta\mu_{B}$ &
$\quad\mu_{B}^{\rm best}$ & $\mu_{\rm exp}$ & $R_B$ \\[0.05in]
\hline\\[-0.1in]
p     &   $\sfrac89 \kappa_u+\sfrac19 \kappa_d+1$ & {\bf 2.793} & {\bf 2.793} &
2.74  & 1.664 & 1.129 &  2.793 &${2.793}$ & 40\% \\[0.05in]
n     &   $-\sfrac29 \kappa_u-\sfrac49 \kappa_d-\sfrac23$ & ${\bf -1.913}$ &${\bf -1.913}$& $-1.96$ & $-1.106$ & $-0.807$ &  $-1.913$  &  ${-1.913}$  & 42\%
\\[0.1in]
\hline\\[-0.1in]
$\Sigma^+$  &  $\sfrac89 \kappa_u+\sfrac19 \kappa_s+\frac{m}{M_\Sigma}$  & 2.530
& 2.674 & 2.58 & 1.971 & 0.491   & 2.462 &  2.45(2)  & 20\%
\\[0.1in]
$\Sigma^0$ & $\sfrac49 \kappa_u-\sfrac29 \kappa_d+\sfrac19 \kappa_s+\sfrac13\frac{m}{M_{\Sigma}}$  & 0.790 &   0.791 & 0.61 & 0.634 & $-0.011$ & 0.623 & --- & $-2$\%
\\[0.05in]
$\Sigma^-$ & $-\sfrac49 \kappa_d+\sfrac19 \kappa_s-\sfrac13\frac{m}{M_{\Sigma}}$  &  $-0.951$ & $-1.092$ & $-1.35$ & $-0.702$ & $-0.513$ &$-1.215$ & $-1.16(3)$ & 42\%
\\[0.1in]
\hline\\[-0.1in]
$\Lambda^0$ &  $-\sfrac13 \kappa_s-\sfrac13\frac{m}{M_{\Lambda}}$  & $-0.768$ &
{\bf $-$0.613} & $-0.57$ & $-0.516$ & $-0.099$  & $-0.615$ &  $-0.613(4)$ &16\%
\\[0.05in]
\hline\\[-0.1in]
$\Xi^0$ &  $-\sfrac29\kappa_u-\sfrac49 \kappa_s-\sfrac23\frac{m}{M_{\Xi}}$ &   $-1.520$ & $-1.435$ & $-1.27$  & $-1.393$ & $0.138$&  $-1.255$ & $-1.250(14)$ & $-$11\%
\\
$\Xi^-$ & $\sfrac19\kappa_d-\sfrac49 \kappa_s-\sfrac13\frac{m}{M_{\Xi}}$  &  $-0.674$  & $-0.493$ & $-0.61$  & $-0.606$ & $-0.141$  &  $-0.747$ &  $-0.65$(3)  & 19\%
\\[0.05in]
\hline
\hline
\end{tabular}
\end{center}
\vspace{-0.1in}
\caption{Results for the magnetic moments of the baryon octet
{\it in nuclear magnetons\/}.
The third column
is the prediction with
{\it no\/} pion cloud, $\kappa_s=1.462$, and $\kappa_u=1.778$,
$\kappa_d=1.915$ fixed to give the correct proton, neutron
and $\Omega^-$ moments \cite{Nucleon,Omega}.
The forth column is the result
of the naive quark model with
nucleon and $\Lambda$ magnetic moments used as input \cite{PDG}.
The fifth 
column is the predictions
of the Cloudy Bag Model \cite{Tsushima
}.
The 6$^{\rm th}$-8$^{\rm th}$ and 10$^{\rm th}$ columns are our predictions ($\mu_B^{\rm best}$)  together with the decomposition    defined in Eq.~(\ref{eq:25})  and the ratio $R_B$
for the dressed moments, with $\alpha$ {\it fixed\/} at 0.6943 and the other coefficients given in the text by
Eqs.~(\ref{eqKfit})-(\ref{eqBCD}).
Column 9 gives the experimental magnetic moments
with their errors \cite{Yao06}.}
\label{tab:4a}
\end{table*}


We now are in a position to write down the
16 equations that describe the
charge (only 4 are independent)
and magnetic moments of the 8 baryons.
We start with the equation for the nucleon charge
\begin{align}
&\sfrac12(1+\tau_3)=Z_N\Big\{\sfrac12(1+\tau_3) + 2B_1\tau_3 + \sfrac12 C_1(3-\tau_3)
\nonumber\\
&\qquad+\sfrac12 D_1(3\kappa_{0N}^s-\tau_3\kappa_{0N}^v)\Big\},
\end{align}
and note that charge conservation requires that
\bea
D_1=0\qquad B_1=C_1. \label{eq:112}
\eea
The CST gives precisely these constraints (as must any model
that satisfies current conservation).
These conditions, which hold for all the baryons in the octet, are the necessary and sufficient conditions that insure the charge of all the baryons is conserved.   The renormalization constant that follows,
\bea
&&Z_N=(1+3B_1)^{-1}
\eea
can also be derived from Eq.~(\ref{eq:3}) [it can be shown that the coefficients multiplying $B_1$ in the renormalization factors $Z_B$ are identical to the $G_{0B}$ derived in Table \ref{tab:2}].   Hence the additional renormalization constants are
\bea
&&Z_\Xi=\Big[1+3\beta_\Xi B_1\Big]^{-1}
\nonumber\\
&&Z_\Sigma=\bigg[1+\Big(2\beta_\Sigma+\beta_\Lambda\Big)B_1\bigg]^{-1}
\nonumber\\
&&Z_\Lambda=\Big[1+3\beta_\Lambda B_1\Big]^{-1}.
\eea

Knowing that the model treats the charges correctly, we now turn to the 8 remaining equations for the magnetic moments.
The equations for the anomalous moments 
(in natural baryon units)
are:
\begin{align}
\kappa_p&=Z_N\Big[\kappa_{0p}+ D_2(\kappa_{0p}+2 \kappa_{0n}) + 2B_2 +C_2\Big]
\nonumber\\
\kappa_n&=Z_N\Big[\kappa_{0n}+D_2(2\kappa_{0p}+\kappa_{on}) -2B_2+2C_2\Big]
\nonumber\\
\kappa_\Lambda&=Z_\Lambda\bigg\{\kappa_{0\Lambda}+\beta_\Lambda D_2 (\kappa_{0\Sigma^+}+\kappa_{0\Sigma^0}+\kappa_{0\Sigma^-}) \bigg\}
\nonumber\\
\kappa_{\Sigma^+}&=Z_\Sigma\bigg\{\kappa_{0\Sigma^+} +D_2\Big[\beta_\Sigma(\kappa_{0\Sigma^+}+\kappa_{0\Sigma^0})+\beta_\Lambda\kappa_{0\Lambda}
\Big]
\nonumber\\&\qquad
+\beta_\Sigma (B_2+C_2) +\beta_\Lambda B_2
\bigg\}
\nonumber\\
\kappa_{\Sigma^0}&=Z_\Sigma\bigg\{\kappa_{0\Sigma^0} +
D_2\Big[\beta_\Sigma(\kappa_{0\Sigma^+}+\kappa_{0\Sigma^-})+\beta_\Lambda\kappa_{0\Lambda} \Big]\bigg\}
\nonumber\\
\kappa_{\Sigma^-}&=Z_\Sigma\bigg\{\kappa_{0\Sigma^-} +D_2\Big[\beta_\Sigma(\kappa_{0\Sigma^0}+\kappa_{0\Sigma^-})+\beta_\Lambda\kappa_{0\Lambda}
\Big]
\nonumber\\&\qquad
-\beta_\Sigma (B_2+C_2)-\beta_\Lambda B_2
\bigg\}
\nonumber\\
\kappa_{\Xi^0}&=Z_\Xi\Big[\kappa_{0\Xi^0}+\beta_\Xi D_2
\Big(\kappa_{0\Xi^0}+2\kappa_{0\Xi^-} \Big)
+ 2 \beta_\Xi (B_2 - C_2) \Big] \nonumber\\
\kappa_{\Xi^-}&=Z_\Xi
\Big[\kappa_{0\Xi^-}+\beta_\Xi D_2
\Big(2 \kappa_{0\Xi^0}+\kappa_{0\Xi^-} \Big)
-\beta_\Xi(2B_2+C_2) \Big],
\label{eq:19}
\end{align}
The charge factor $e_B$ must be added to
each of these to get the total magnetic moment, and to convert to nuclear magnetons each equation is multiplied by $m/M_B$.  The ''bare'' magnetic moments, assembled from the quark moments as shown Eq.~(\ref{eq:7}), are tabulated in Table \ref{tab:4a}.

The eight Eqs.~(\ref{eq:19}) and the results for the
bare magnetic moments in terms of the quark anomalous moments and the quark charges,
found in Table \ref{tab:4a},  give the eight baryon magnetic moments (seven of which are measured)
in terms of six parameters:  the quark anomalous moments $\kappa_u$ and $\kappa_d$ and the four pion cloud constants $B_1, B_2, C_2, D_2$
($\alpha=0.6943$ was determined from the self energies and we use the strange quark anomalous moment, $\kappa_s$,
fixed previously from our study of the
$\Omega^-$ magnetic moment which has no pion cloud  \cite{Omega}).

Our best fit is shown in Table \ref{tab:4a}.   To obtain this fit we minimized $\chi^2$ using the experimental errors listed, except for the proton and neutron, which were assigned an experimental error of 0.0001, about 1000 times larger than the actual errors.  Only two of the magnetic moments are significantly  outside the (redefined for the nucleons) experimental errors:  the $\Sigma^-$ (about 2 standard deviations) and $\Xi^-$ (about 3 standard deviations) and these two have the largest experimental errors.    For the anomalous magnetic moments
we obtain
\be
\kappa_u= 1.929 \qquad \kappa_d= 1.911.
\label{eqKfit}
\ee
Compared to Ref.~\cite{Nucleon}, the pion cloud contributions increased  $\kappa_u$ by about 8\%,
bringing it very close to $\kappa_d$ (which is almost unchanged).
The pion cloud coefficients determined by the fit are:
\bea
&B_1=0.2531\qquad &B_2=0.5648\nonumber\\
&C_2=-0.06599\qquad &D_2=-0.08321\, .
\label{eqBCD}
\eea
The main point of this letter is that our fit has {\it constrained\/} the size and structure of the pion cloud contributions to the octet moments, and hence the {\it size of pion cloud contributions to the nucleon form factors\/}.

Table  \ref{tab:4a} also compares our best fit to three other models, two without a pion cloud.   The predictions of the ``bare'' moments (given analytically in the second column) are fit with two parameters ($\kappa_u$ and $\kappa_d$) and have a maximum deviation of 0.27$\mu_N$; the naive quark model reported in Ref.~\cite{PDG} has a maximum deviation of $0.22 \mu_N$, and the cloudy bag calculation of Ref.~\cite{Tsushima} has a maximum deviation of $0.19\mu_N$.  Our maximum deviation is 0.10$\mu_N$.

The description of the ratio $r=\mu_{\Xi^-}/\mu_{\Lambda}$ has been a longstanding problem for
SU(6) constituent quark models, where $r<1$ if  $m_u < m_s$.  Models without a pion cloud, such as those of  Refs.~\cite{Nucleon,Omega} and the NQM
\cite{PDG}
,  suffer from this limitation.  The experimental value is $r \simeq 1.06>1$, and both the
cloudy bag model \cite{Tsushima} and our new results give $r>1$.

\begin{table}
\begin{center}
\begin{tabular}{l c c }
 &    $\mu_{p}$ & $\mu_{n}$  \\
\hline
$\chi$PT \cite{Puglia00} &  --1.901 & 1.291 \\
$\chi$PT \cite{Geng09}  &  --0.85  &  0.63 \\
CBM \cite{Leinweber99}  &    1.5 &  --0.9 \\
Barik \cite{Barik86}      &  0.43  & --0.44 \\
Ha \cite{Ha98}           &  --0.008 & --0.128 \\
Faessler \cite{Faessler06}  &        0.44 & --0.34  \\
Miller   \cite{Miller02}  &          0.32 &--0.38 \\
Wang \cite{Wang07}        &          0.26 & --0.41 \\
Cloet \cite{Cloet08}       &         0.24 & --0.40   \\[0.1in]
This work &  1.129 &$-0.807$\\
\hline
\end{tabular}
\end{center}
\caption{Model calculations of pion cloud contributions to the nucleon magnetic moments.
}
\label{tab:7}
\end{table}

To give a measure of the size of the pion cloud contributions to each baryon, we decompose the magnetic moment into two terms
\be
\mu_B^{\rm best}= Z_B \left[\mu_{0B} + \delta \mu_B \right] \, . \label{eq:25}
\ee
Table \ref{tab:4a} shows the separate contributions $Z_B\,\mu_{0B}$ and $Z_B\,\delta\mu_B$, as well as the ratio $R_B=Z_B\,\delta\mu_B/\mu_{B}^{\rm best}$.  The pion cloud corrections to the nucleon and $\Sigma^-$ moments (as measured by $R_B$) are all about 40\%, with the corrections to the other moments much smaller.  For comparison, Table \ref{tab:7} shows the size of pion cloud contributions for a number of recent models of the nucleon form factors.  Our results for the neutron correction are close the the Cloudy Bag Model of Ref.~\cite{Leinweber99},  but  our proton correction is almost 40\% smaller.  The other calculations are either too small or of  the opposite sign.

We close this letter with a brief comment about the fit to the magnetic moments and the reliability of our estimate  (\ref{eqBCD}) of the pion cloud coefficients.  It is natural to ask (as we did) why we cannot obtain a better fit to the seven known magnetic moments with six parameters.  To study this we allowed the seventh ($\alpha$) to vary from the value determined by the fits to the baryon masses (0.6943).  We found three minima at $\alpha=0.2523, 0.6948,$ and 1.3153.  The minima at 0.25 and 1.3 both gave essentially perfect fits to all of the magnetic moments, with pion cloud coefficients several times larger (at $\alpha\simeq0.25$) or very small (at $\alpha\simeq1.3$).  The other minimum, at {\it almost the same point\/} determined by our fit to the masses, gives the (not so perfect) fit shown in Table \ref{tab:4a}.   We conclude that (i) the determination of the coefficients (\ref{eqBCD}) is strongly dependent on the physical constraint that $\alpha$ be near 0.6, as favored by SU(6) and our fit to the baryon octet masses, and (ii) the {\it same\/} value of $\alpha$ (almost) provides both the best fit to the masses and the best (local) fit to the magnetic moments.


\vspace{1ex}
\noindent
{\bf Acknowledgments:}

This work was partially support by Jefferson Science Associates,
LLC under U.S.~DOE Contract No.~DE-AC05-06OR23177.
G.~R.\ was supported by the Portuguese Funda\c{c}\~ao para
a Ci\^encia e Tecnologia (FCT) under the grant
No.~SFRH/BPD/26886/2006.
This work has also been supported in part by the European Union
(HadronPhysics2 project ``Study of strongly interacting matter'').



\end{document}